\date{}
\documentstyle[preprint,eqsecnum,aps,epsf]{revtex}
\begin{document}
\title
{Persistent currents in mesoscopic cavities and the effect of level crossings
as random variables}
\author{A. J. Fendrik and M. J. S\'anchez}
\maketitle
\noindent
\begin{center}
\center{\it Departamento de F\'{\i}sica  J. J. Giambiagi,\\
 Facultad de Ciencias Exactas  y Naturales, \\
Universidad de Buenos Aires.
Ciudad Universitaria, 1428 Buenos Aires, Argentina.}\\ 
\end{center}
\begin{abstract}
In  the present article we  perform analytical and numerical calculations  related  to persistent currents in 2D isolated mesoscopic annular cavities threaded  by a magnetic flux.
The system considered  has a high number of  open channels and therefore the single particle spectrum exhibits many level crossings as the flux varies.
We determine the effect of the distribution of level crossings in the 
typical persistent current.

\end{abstract}

\section*{Introduction}
\label{int}
Recent developments in micrometer-scale technology have made possible the fabrication of devices small enough to make
electrons  phase-coherent inside. Therefore the transport properties of those samples at very low temperatures  exhibit features characteristic of quantum coherence of the electronic wave function along the sample.
The observed Aharonov-Bohm oscillations in the resistance of a loop pierced by  a magnetic flux is one of the most relevant manifestations of the interference phenomena typical of phase coherence \cite{but,was}.\\
Although the transport properties have been intensively investigated both theoretical and experimentally, the persistent current problem remains less understood. In 1983 B\"uttiker, Imry and Landauer, following earlier work by Byers and Yang on superconducting rings \cite{byers}, proposed the existence of such currents in one dimensional mesoscopic loops in the presence of weak magnetic fields \cite{but2}. These  equilibrium currents are a consequence of the nature of the eigenfunctions and their flux sensitivity, which is strictly of the Aharonov-Bohm type. The current  is a periodic function of the magnetic flux  with  fundamental period $\phi_{0}= h c /e$. In ideal clean rings at 0 K  ({\bf ballistic regime}) the current  is proportional to  $I_{0} \equiv e \; v_{F}/{L}$ where $v_{F}$ is the Fermi velocity and $L$ is the length of the loop. On the other hand, in the diffusive regime (when impurities are present
and the elastic mean free path {\it l} is shorter than the typical sample size)  theoretical  studies predict that the current decreases by a factor ${\it l} / L$.
The problem  of persistent currents did not deserve much attention until  recent years when experimental measurements  have been performed  on metallic rings in the diffusive regime \cite{levy,chan} and in a few-channel 
mesoscopic semiconductor ring \cite{mailly}. For the metallic systems, the experimental value of the current is $\sim (0.3-2) I_{0}$, 1 or 2 orders of magnitude greater than the theoretical predictions consistent with diffusive motion, that is with no apparent corrections due to disorder. The value of the current reported by  Mailly and collaborators on the mesoscopic ring,  is also of the order of $I_{0}$, but in their experiment ${\it l}/L \sim 1$. \\
Owing to the discrepancies between the observed and predicted values of the current, much of the theoretical efforts were put to  find a mechanism which could compensate the effect of impurities. We are not going to describe those approachs here. For a recent review of the problem, see  \cite{guhr}. \\
Most of the theoretical studies mentioned above where restricted to one dimensional geometries \cite{cheung1}. 
The effect of the number of channels in the persistent current, although
investigated by many authors, is not yet completely understood. The computations of these currents in 2D (multichannel) geometries is much more difficult, and all the  studies for multichannel geometries have been performed employing discrete models or  cylindrical geometries in which the transverse channels are decoupled for the conduction mode \cite{cheung2,bouch,willy}.\\

In this paper we deal  with  the persistent current problem 
for a system of $N$ non-interacting electrons confined 
in a  mesoscopic clean annular 2D geometry  at 0 K. This geometry is useful to describe  the real metallic ``rings'' employed in most of the experiments mentioned above. As an example in Ref.\cite{mailly} the ring has internal diameter $2 \mu m$ and external diameter $3.4 \mu m$, conforming a 2D cavity.

Our system is formally a quantum billiard whose hamiltonian $H$ depends on a
parameter (the magnetic flux). The variation of the parameter
preserves the commutator $[H,L]=0$ ($L$ being the angular momentum), and therefore the single particle spectrum displays 
large amount  of crossings as the parameter is varied \cite{fendrik}. 
Such degeneracies and their features are fundamental ingredients that affect
the persistent currents when many channels are open. In fact,
they are not only responsible for the multiple jumps that the persistent 
currents show as a function of flux (for fixed number of 
particles $N$) but also for the large oscillations in the typical current as a function of $N$ after averaging on the flux. 
These  oscillations (which are inherent to the system because they depend on the features of the crossings of  the single particle energy levels) make
difficult to establish an average behavior with $N$. 
In the present work we show that not only the actual average behavior 
but also the large oscillations emerge when the statistical properties of the
degeneracies are considered.\\
The paper is organized a follows.
In Sec.~\ref{ab}, we present our system (we call it the Aharonov-Bohm annular 
billiard). Sec.~\ref{pc} is devoted to summarize some results and properties
of persistent currents in 1D and 2D. In Sec.~\ref{dc}, we introduce the
properties of level crossings that allow us to perform a statistical approach
to the problem. Section~\ref{tc} is addressed to exploit the properties of
the level crossings in order to determine the dependence of the average behavior and the fluctuation  of the persistent current on the number of particles $N$. Finally, in Sec.~\ref{cr} we present the concluding remarks.

\section{The Aharonov-Bohm annular billiard}
\label{ab}
\begin{figure}
\begin{center}
\leavevmode
\epsfysize=10cm
\epsfbox{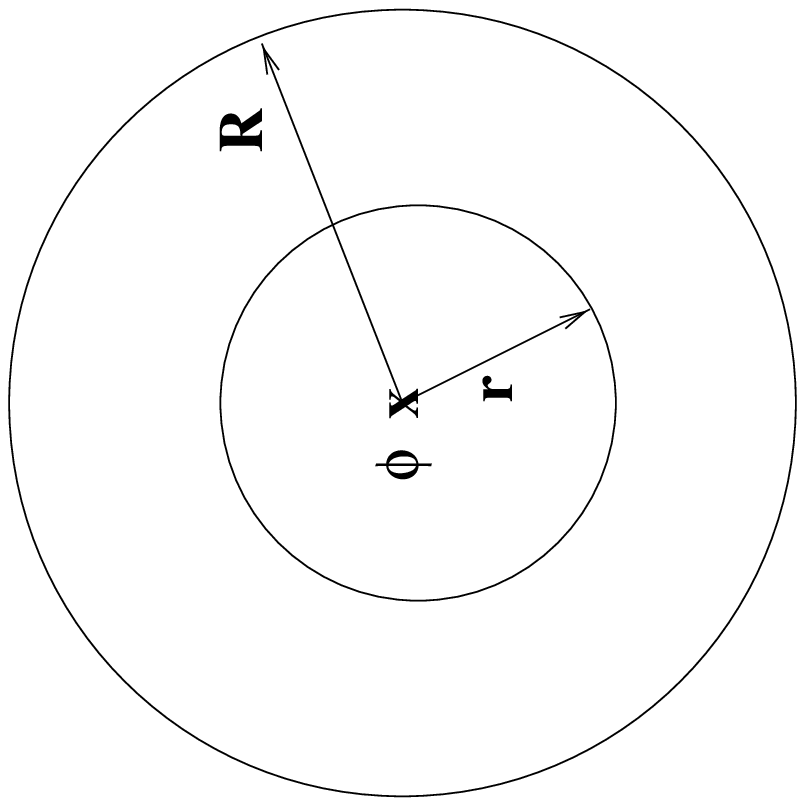}
\end{center}
\caption{Annular billiard threaded by a magnetic flux $\phi$. The internal and external radii are denoted $r$ and $R$ respectively.} 
\label{figure1}
\end{figure}

We consider a 2D  structure with the geometry of an annulus. See 
Fig.~\ref{figure1}.
The cavity consists on the planar region limited by two concentric circles of 
outer and inner radii $R$ and $r$ respectively. In the following we will take the area equal to $\pi$ and we define the parameter $\lambda ={ R / r}$. The spatial degrees of freedom are the azimutal angle $\theta$ and the radial coordinate $\rho$ which varies between $r$ and $R$.\\
Let N be the number of non-interacting electrons on the cavity.
We are interested in the persistent current caused by the application of an homogeneous time independent magnetic flux $\phi$ threading the cavity axially.
We disregard the effects of the magnetic field piercing the body of the annulus, so the electrons move with uniform rectilinear motion inside the cavity. 
We choose the gauge in which $\vec{A}= \phi/(2 \pi \rho) \; \hat{\theta}$
where $\hat{\theta}$ is the azimutal unit vector.

The single particle spectrum results from the eigenvalue equation
\begin{equation} 
\label{he}
\Delta  \Psi   + \frac{2 \; i \;  {\alpha}}{\rho ^{2}} \frac{\partial  \Psi}{\partial \theta} - \frac{\alpha ^2}{ \rho ^2} \Psi + k^2 \Psi  =  0
\end{equation}
where $\Delta$ is the Laplacian in polar coordinates. We define  the scaled flux $\alpha= \phi / \phi_{0}$ and  we use units such ${\hbar^2}/ 2 m =1$, so the energy is  $E = k^2$.

We apply  Dirichlet boundary conditions at $\rho= r$ and $\rho= R$ and periodic boundary condition in the azimutal direction. 

 The Eq.(\ref{he}) is separable in polar coordinates and we    
factorize $\Psi (\rho, \theta)= {\cal  F}{(\rho)} \; \exp{i m \theta} \;$ with 
$m=0,\pm 1,\pm2...$ the orbital quantum number. The eigenfrequencies $k_{\nu,n}$  results from the solution of the equation 
\begin{eqnarray}
\label{ceros}
J_{\nu }(z \lambda) N_{\nu}(z) - J_{\nu } (z) N_{\nu}(z \lambda)  =  0 \; , \nonumber \\
\nu  =   m + \alpha \; ,
\end{eqnarray}
where we have defined $z \equiv  k r $ and $n =1,2...,$ is the radial quantum number.  
$J_{\nu}$ and  $N_{\nu}$ are the Bessel functions of the first and second kind, respectively.
The corresponding eigenfunctions $\Psi(\rho,\theta)$ are :

\begin{equation}
\Psi_{\nu,n}(\rho, \theta)= A_{\nu,n} [ J_{\nu } (k_{\nu,n} \rho) N_{\nu}(k_{\nu,n} r) - J_{\nu } (k_{\nu,n} r) N_{\nu}(k_{\nu,n} \rho) ]\; 
\exp (i m \theta) \;
\end{equation}
where $A_{\nu,n}$ is the normalization constant.\\
All the eigenstates and all the equilibrium 
physical properties of the system are periodic functions of the flux with period $\phi_{0}$ \cite{byers}. Moreover as the energy spectrum is symmetric with respect to $\phi= \phi_{0} / 2$,  in the following the parameter $\alpha$ will take values between $0$ and $1/2$.
For $\alpha\ne 0$, the states with $m$ and $-m$
are, in general, not degenerate. This is the reason for the existence of an equilibrium persistent current.\\

We stress that for the present system   $E_{\nu,n}$ can not be written down as a simple function of  the numbers $\nu$ and $n$ as it happens, for example, in the  cylindrical geometries in which the eigenenergies are cuadratic functions of the quantum numbers. This constitutes the fundamental difference between the annular cavity and  other 2D geometries studied so far, \cite{cheung1}.

\begin{figure}
\begin{center}
\leavevmode
\epsfbox[88 35 575 524]{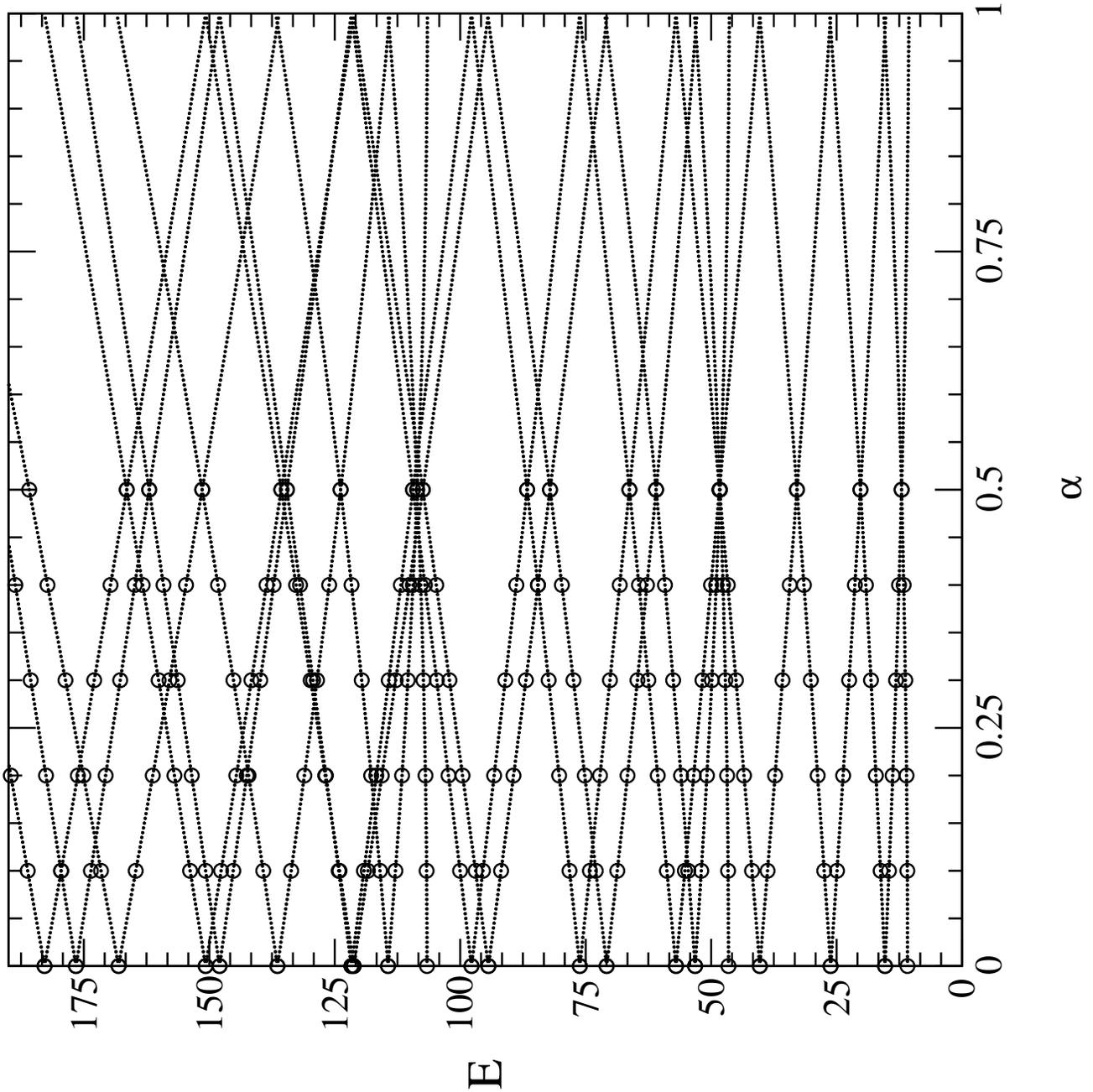}
\end{center}
\caption{Single particle energy levels as a function of the adimensional flux $\alpha$, corresponding to the lowest region of the spectrum for the annular billiard with $\lambda = 10$. The circles correspond to exact values  obtained  from the zeros of the cross products of Bessel functions. The small dots lines are the quadratic fits. See the text for details.} 
\label{figure2}
\end{figure}

To obtain the eigenenergies $E_{\nu,n}$  we have numerically solved Eq.~(\ref{ceros}).
For each fixed value of $\lambda$, $m$ and $n$, we have taken  six equally spaced values of $\alpha$ between $0$ and $1/2$. Then we have verified that  the best quadratic fit was quite satisfactory for the considered values of $\lambda$.
Fig.~\ref{figure2} shows a region  of the energy spectrum  as a function of  
$\alpha$ for  $\lambda = 10$ obtained by the described procedure. In the following we will consider

\begin{equation}
\label{fiteo}
E_{m + \alpha, n} = A_{m,n}(\lambda) \alpha^2 + B_{m,n}(\lambda) \alpha + 
C_{m,n}(\lambda) \; .
\end{equation}

Let us  remark that for  the cylindrical geometries ({\it ie.} a cylindrical surface of area $L \times L_{y}$) the coefficients can be 
determined exactly and they are, $A= (2 \pi /L)^2$ (without any dependence  on the quantum numbers) and   $B= 2 \; m \; (2 \pi /L )^2 $  which depends only on the orbital quantum number $m$. The other quantum number appears only in the 
constant term $C$.

\section{Persistent currents in N electron  systems}
\label{pc}
In the present section we summarize the main results concerning  persistent
currents in 1D (rings) and 2D (cylinders and annula) systems.\\
For a system with $N$ non-interacting electrons, the total current is $I_{T}= \sum_{n=1}^{N} I_{n}$ \cite{cheung1}.\\
In order to characterize the typical current we define $
I_{typ} = \sqrt {\langle {I_{T}^2} \rangle_{\alpha} }\;$, 
where the symbol $ \langle .. \rangle_{\alpha} $ means flux average. This is
\begin{equation}
\label{itypi}
I_{typ}^2 = 2 \int_{0}^{1/2} I_{T}^2 \; d\alpha \; ;
\end{equation}
where we have profited the symmetry of the spectrum with respect to $\alpha =\frac{1}{2}$.

\subsection{1D Systems}
 
For a 1D ring  of circumference $L$ threaded by a magnetic flux $\phi$,  the velocity of an electron in the state with  orbital quantum number $m$ is $v_{m}= {\frac{1}{\hbar}}\partial{E_{m}}/\partial{k_{m}}$.
Therefore as $k_{m}= 2 \pi \; (m + \alpha)/ L$,
the current carried by the level $E_{m}$  results \cite{cheung1}
\begin{equation}
\label{c1d}
I_{m}= - \frac{\partial{E_{m}}}{ \partial {\phi}} \; .
\end{equation}

It is well known that for the  1D ring geometries, the crossings between 
 levels in the range $0 \le \alpha \le 0.5 $ occur only for $\alpha = 0$  or  
$\alpha = 0.5$. This fact is a direct consequence of the existence of only one channel in the system. The total current for the N-electron system is,
 
\begin{equation}
\label{ct1d}
I_{T}= \left\{ \begin{array}{lll}
        \frac{- 2 \;e} {L^2} \pi \; N \; \alpha     & \; \; \; \; \mbox{$N \; odd$}\\
        \frac{- e }{L^2} \pi  \; N \; (2\; \alpha -1)  &  \; \;  \mbox{$N \; even$}
                         \end{array}
                 \right.  \nonumber \; ,
\end{equation}
where we have explicitely used $k_{F}= \pi N / L$. 
Therefore the typical current is $I_{typ} \sim N$. This is consistent with the well known result for 1D geometries, $I_{typ}= I_{0}$ with $I_{0} \equiv e \; v_{F} / L$ \cite{cheung1}.\\

\subsection{2D Systems}
\label{2d}

In the case of 2D cylindric geometries, as the energy of an individual  state is  separable in two terms, one associated with the conduction mode and the other with the transversal one, it is straigthforward to verify  
 Eq.(\ref{c1d}).
For the annular geometry, that equation is not obvious. In order  to prove   Eq.(\ref{c1d})  we begin computing  the current density     $\vec { j}$ : 
\begin{equation}
\label{j} 
 \vec { j}  = \frac{1}{ 2 \; m} [\Psi^* \;( -i \; \hbar  \nabla  + \frac{e}{c}  \vec A )\; \Psi   + \Psi \; ( i \; \hbar  \nabla   + \frac{e}{c}  \vec A )\; \Psi^*   ] \; .
\end{equation}
 
From the Eq.(\ref{he})  it is straightforward to show that 
\begin{equation}
\label{dh}
{\frac{\partial{\cal  H}}{\partial \phi}} = \frac{e }{ m \; c}[ -i \; \hbar  \nabla + \frac{e}{c}  \vec A ] . \; \frac{\partial \vec A}{\partial\phi} \; .
\end{equation}
Comparing  Eq.(\ref{dh}) and  Eq.(\ref{j}), we obtain
\[
\frac{2 \; e}{c} \; \vec { j} \; . \frac{\partial\vec A}{\partial\phi} =  \Psi^*{\frac{\partial{\cal  H}}{\partial \phi}} \Psi \; + \; \Psi \; \left( {{\frac{\partial{\cal  H}}{\partial \phi}}} \right)^* \Psi^* \; .
\]
Taking into account the functional form of the vector potential $\vec A$ and   the Hellmann-Feynman theorem  \cite{mer}, the current carried by the state $i$ results:
\begin{equation}
\label{c2d}
I_{i} \equiv \; - \; e \int j_{\theta} \; d\rho = \; - c \; {\frac{\partial E_{i}}{\partial\phi}} \; ,
\end{equation}
where  $E_{i} \equiv {k_{i}}^2$ and the  limits of integration are $r$ and $R$  respectively. \\
Therefore, following  Eq.(\ref{c2d}), the total persistent current $I_{T}$, is the current through  a tranverse section of the annulus.
 
For 2D geometries, as we have already mentioned, the effect of the number of channels on the typical currents   is not yet fully understood. The number of channels $M$ is defined as the maximun value of the transverse quantum number  inside the Fermi surface at zero flux. We define the value of  Fermi energy  at zero flux as $E_{F}\equiv
E_{N}$, $E_{N}$ being the energy of the highest ocuppied level. We  emphasize that in the considered isolated cavities  the number of electrons $N$ is constant for all values of the flux. For values of $\alpha \neq 0$,  the average Fermi energy is defined by Weyl's formula $\langle E_{F} \rangle = N / 4$ (in the stipulated units) \cite{bal}.\\ 
One important point is to know how $M$ depends on  the number of particles  $N$ and on the  parameter $\lambda$.
Employing the expansion of the zeros of the radial functions (solutions of the Eq.(\ref{ceros})) we obtain \cite{abram}:

\[
M^2\; = \left[ 1 + \frac{ 4 \; N \; {\left( \lambda  -1 \right)} ^2}{{\pi }^2  \; ({\lambda}^2 -1)} + {\left( \frac{{\left( \lambda  -1 \right)} ^2}{8 \pi^2 \; \lambda}\right)}^2 \; \right]\; ,
\]
where the symbol $[..]$ means integer part. \\
For $\lambda \rightarrow 1$, the second and third terms in the last equation go to 0 and  we obtain the trivial result  for 1D geometries,  $M=1$ independently  of the number of particles $N$.

Most of the  experimental relevant  situations involve many particles. For  values of $\lambda > 1 $ and  for $N >> 1$  the second term  in the last expression dominates and $M$ reads: 
\begin{equation}
\label{efs}
M  = \left[ \frac{2 \; \sqrt{N} \;(\lambda -1)}{ \pi \; \sqrt{\lambda ^2 -1} } \right] \; .
\end{equation}

\begin{figure}
\begin{center}
\leavevmode
\epsfysize=10cm
\epsfbox[84 33 576 524]{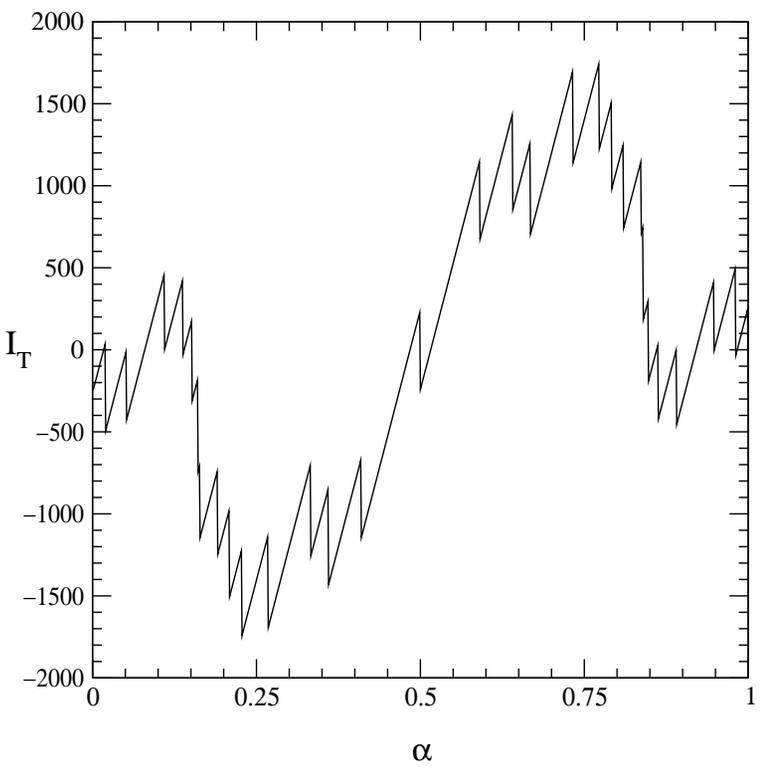}
\end{center}
\caption{Persistent current $I_{T}$ as a function of the adimensional flux $\alpha$ for $N=2474$ and $\lambda=10$. } 
\label{figure3}
\end{figure}

Fig.~\ref{figure3} shows the total current vs flux for $N=2747$ and $\lambda = 10$.
This corresponds to $ M = 60$. The existence of many channels implies that, opposite to what happens for 1D geometries, crossings between levels of different channels occur  for $0 < \alpha < 0.5 $.\\
Employing the expansion given by Eq.(\ref{fiteo}), the total current for the $N$ electrons system can be written as :
\begin{equation}
\label{ct}
I_{T}= \alpha \; ( \sum 2 \; A_{m,n}) +  \sum  B_{m,n} \; ,  	
\end{equation}
where the coefficients $A's > 0$, and the coefficients $B's$ have alternated signs. The sum extends to all the filled levels $N$.
The total current has two contributions, the first term which has an explicit linear dependence on $\alpha$ and the second one without explicit dependence on the flux. Nevertheless, Fig.~\ref{figure3} shows  the total current which looks like a sawtooth. Between two succesive crossings the curve is a linear function with  the same positive slope in each interval. The negative jumps in $I_{T}$ are for values of the flux  {\bf $\alpha_{i}^{(-)}$} where crossings  between states occurs. This qualitative behavior can be understood as follows.
Let us suppose that a given state is occupied before a crossing and empty after it. This corresponds to replace in the Eq.(\ref{ct}) one value of $A_{m,n}$ by another one. As $N >>1$, the last replacement gives essentially the same slope.
On the other hand, the second term is strongly affected because a positive value
of $B_{n,m}$ is replaced by a negative one, and this produces a significative
{\bf negative} change in the value of the non homogeneous term of $I_{T}$.
Let us remark that between two adjacent $\alpha_{i}^{(-)}$ 's there is  another  crossing at  ${\bf \alpha_{i}^{(+)}}$ that is irrelevant for $I_{T}$.
We denote $n_{c}^{(+)} (n_{c}^{(-)})$ the total number of crossings at the positions $\alpha_{i}^{(+)} (\alpha_{i}^{(-)})$.
As it will be clarified in the following sections, the features of the crossings between levels are strongly related to the shape and the magnitude of the total and typical currents.
  
\section{The distribution of crossings}
\label{dc}
\begin{figure}
\begin{center}
\leavevmode
\epsfysize=10cm
\epsfbox[84 33 578 538]{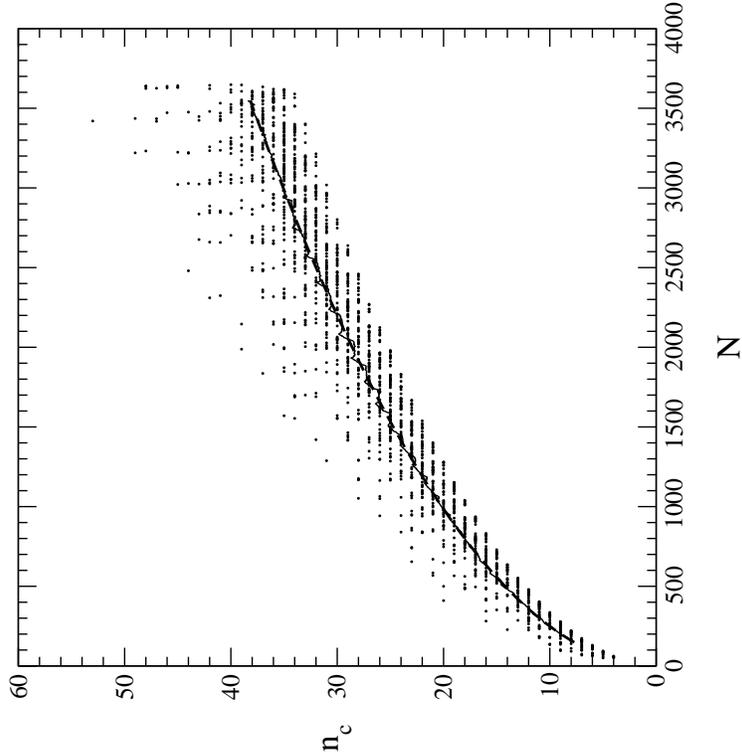}
\end{center}
\caption{Number of crossings $n_{c}$ for $0<\alpha < 1/2$ as a function of the number of levels $N$. The points are the exact values numerically calculated. The thin line corresponds to an average over $\Delta N=100$ while the dashed line corresponds to the best power fit ($n_{c}=0.61822 \times N^{0.50441}$) to this average.}  
\label{figure4}
\end{figure}

\begin{figure}
\begin{center}
\leavevmode
\epsfysize=10cm
\epsfbox[29 24 588 586]{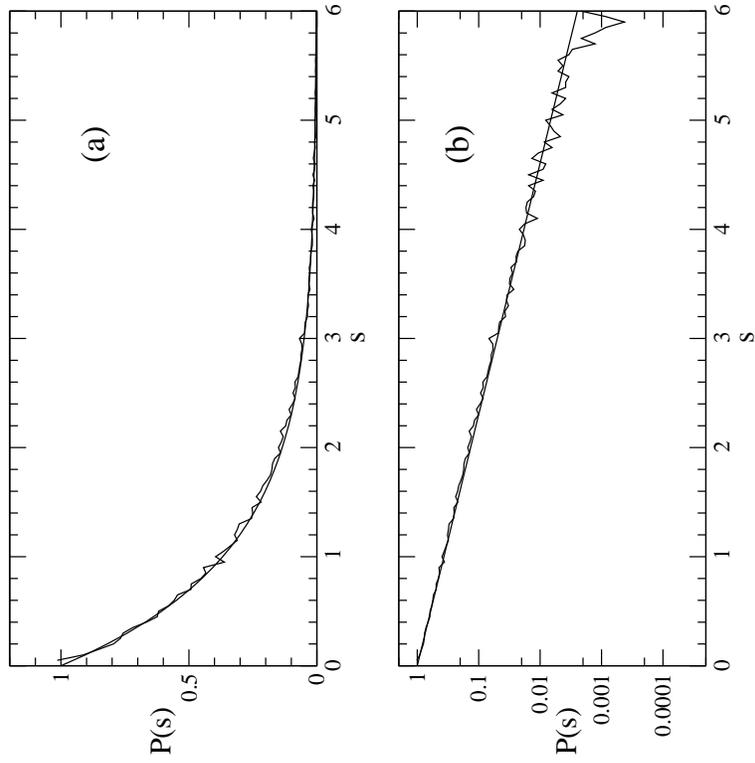}
\end{center}
\caption{ (a) Nearest neighbour spacing distribution $P(s)$ numerically calculated. We also show the exact Poisson distribution $P(s)=\exp{-s}$. (b) Log-lin plot of $P(s)$.}  
\label{figure5}
\end{figure}

We devote  this section to establish some relevant properties of the crossings between states for  integrable  hamiltonian systems that depend on a single parameter.
Following the semiclassical expansion for the density of states developed by Berry and Tabor for integrable systems \cite{berry}, 
 we have started writting the density of crossings per unit of energy and flux (for a given value of the parameter $\lambda$) as
\[
\rho_{c}(E,\alpha,\lambda) = \tilde{\rho}_{c} + \; {\rho_{c}}^{osc} \;;
\]
 $\tilde{\rho}_{c}$ is the average  density of crossings (where the average is taken over energy and flux) and   ${\rho_{c}}^{osc}$ is the fluctuating part. 
We have demonstrated that
\begin{equation}
\label{cruces}
 \tilde{\rho}_{c} = C(\lambda) \sqrt{n} \; ,
\end{equation}
 and is independent of $\alpha$ \cite{fs2}. Therefore the average number of crossings
that a given level $n$ has between $0 < \alpha < 0.5 $ is 
$\tilde {n}_{c} = 0.5 \; \tilde {\rho}_{c}$. 
Moreover,
\begin{equation}
\label{cruces2}
 \tilde{n}_{c}^{(+)} = \; \tilde{n}_{c}^{(-)} = \; \tilde{n}_{c}/2 \; .
\end{equation}
Fig.~\ref{figure4} shows  $n_{c}$ vs. $n$, $n_{c}$ being the number of crossings of the level $n$,  
for $\lambda = 10$ and $N= 3650$, together with the ensemble average $\bar{n}_{c}$ taken on a range $1 << \Delta n < N$.
The best   power fit to $\bar{n}_{c}$ is
$\bar{n}_{c}= 0.6182 \; {n}^{\beta}$ with $\beta = 0.50441$, in strong agreement with our theoretical prediction. 
It is  interesting to note that while the number of channels scales as $M \sim \sqrt{N}$ (see Eq.(\ref{efs})),
the mean number of crossings for the highest occupied level  is proportional to $M$. That means that on average, the last occupied level experiments as many crossings as there are channels present in the system.
The higher fluctuations observed in the Fig.~\ref{figure4} correspond
to the apparition of a state with $m=0$ (at $\alpha=0$) in the energy region considered. \\
Another important result concerns  the way in which the crossings for a given level $n$ are distributed in the interval $0 < \alpha < 0.5$. We define the normalized nearest neighbour spacing as $s_{i}= {\tilde n}_{c} \,({\alpha_{i} - \alpha_{i-1}})$, where $\alpha_{i}$ denotes the position of the crossing $i$. 
Fig.~\ref{figure5} shows the nearest neighbour spacing distribution $P(s)$, obtained after an exhaustive numerical computation. This results in a Poisson distribution as can be checked in the mentioned figure,
where the numerical result  is displayed  along with the exact Poisson distribution. In other words, for a given level $n$ with $n_{c}$ crossings, the  corresponding $\alpha_{i}'s$ can be seen as a sample of $n_{c}$  uncorrelated random variables in the interval $0 < \alpha < 0.5$. 
Therefore, if we generalize the above definitions to n-nearest neighbour spacings as  $s_{i}= {\tilde n}_{c} \,({\alpha_{i+n} - \alpha_{i}})$,  the associated spacing distributions are \cite{metha},
\begin{equation}
\label{poison}
P(n,s) = \frac{s^{n}}{n!} \; \exp{(- s)} \; .    
\end{equation}
In the notation of the last equation $P(s) \equiv P(0,s)$. Such distributions of the spacings determine the fluctuations of the typical current, as we will show in the forthcoming section.\\

\section{Typical currents of 2D systems and the crossings as random variables}
\label{tc}

\begin{figure}
\begin{center}
\leavevmode
\epsfysize=10cm
\epsfbox[19 22 589 569]{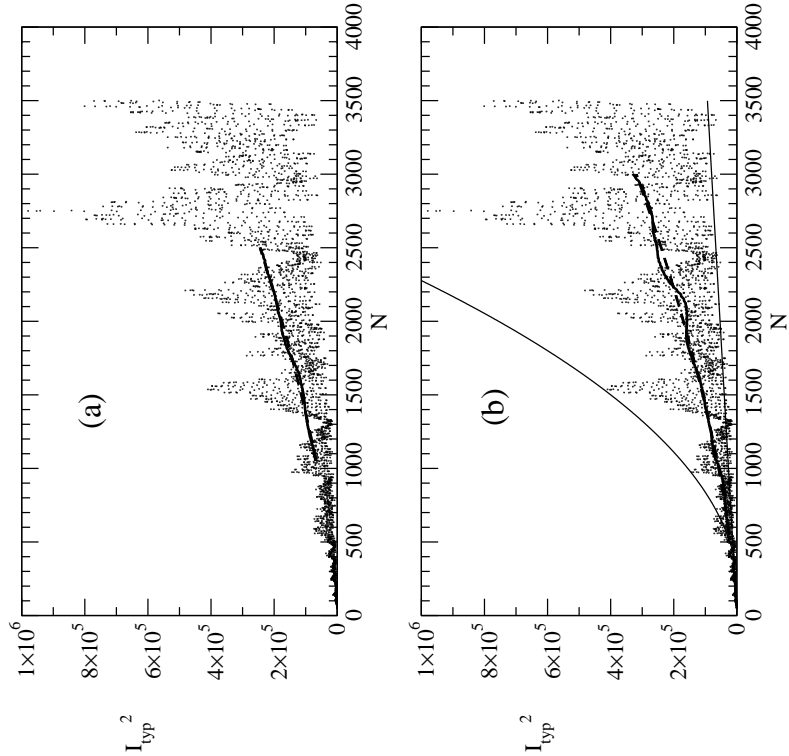}
\end{center}
\caption{(a) The square of the typical current $I_{typ}^2$  as a function of $N$. The points correspond to exact values numerically calculated. The solid line corresponds to an $N$ - average with $\Delta N=2000$, while the dashed line is the best power fit $I_{typ}^2 = \; 2.009 \times N^{1.4966}$. (b) The same exact results of a) but the $N$-average was performed with $\Delta N=1000$. In this case the best power fit is $I_{typ}^2 = \; 1.7405 \times N^{1.5558}$. Here, we have drawn the limit curves ($\propto  N^2$ and $\propto  N$)  for particular realizations of sequences $\alpha_{i} 's$.} 
\label{figure6}
\end{figure}

At this point  we have  all the ingredients to quantify the way in which the distribution of crossings determines the scaling  of the typical current with the number of particles present in the system $N$.\\
We want to  compute $I_{typ}^2$ (see Eq.~(\ref{itypi})).  
Taking into account the definition of the total current Eq.(~\ref{ct}), and the comments stated at the end of Sec.\ref{2d}, we can write the square of the total current  for $ 0 < \alpha < 0.5$ as, 
\begin{eqnarray}
\label{itt}
I_{T}^2 & = & \left( A_{T} \; \alpha  - F(\alpha ,\alpha_{i}^{(-)}) \right)^2  \; , \\
F(\alpha,\alpha_{i}^{(-)}) & \equiv &\sum_{i=1}^{n_c^{(-)}} \Delta_{i} \; \Theta (\alpha \; - \alpha_{i}^{(-)}) \; ,
\end{eqnarray}
where we have defined $A_{T}\equiv 2 \; \sum A_{m, n}$ (see Eq.(\ref{ct})) and   
$\Theta (x)$ is the step function.  
The   positive definite quantity  $\Delta_{i}$ is the  absolute value of the jump that the total current has at the position $\alpha_{i}^{(-)}$. As we  mentioned before, all the jumps have the same negative sign, therefore we have written down a  minus sign preceding the  second term of Eq.~(\ref{itt}). 
We stress that in the last equation $n_{c}^{(-)}$ denotes the total number of crossings (for $ 0 < \alpha < 0.5$)  for which $I_{T}$ shows jumps. 
For the sake of symplicity in the notation, in the following we drop the supraindex in the $\alpha_{i}^{(-)}$ and in $n_{c}^{(-)}$.
$I_{typ}$, and obviously its squared value, can be  written as a function of the random variables  $\alpha_{i}$,
\begin{equation}
\label{itypi2}
{I_{typ}}^2 = 2\; \left( \frac{A_{T}^2}{24} +\;  \sum_{j,i=1}^{n_c} \Delta_{i} \;\Delta_{j}\left(\frac{1}{2} \; - Max(\alpha_{j},\alpha_{j})\right) - \; A_{T} \; \sum_{i=1}^{n_c} \Delta_{i} \;\left(\frac{1}{4} \; - \alpha_{i}^2\right) \right) \; , 
\end{equation}
where $Max(a,b)$ means the maximum among $a$ and $b$. Figure ~\ref{figure6}
shows the exact numerically obtained  $I_{typ}^2$  as a function of the number of particles in the system $N$ for $\lambda = 10$. The extremely fluctuating behavior with $N$ is displayed.\\ 
Our aim is to determine the average dependence of ${I_{typ}}^2$ with the total number of particles in the system $N$. This  implies  performing an  average over $N$ when  the number of particles varies in a range $1<< \delta N < N$ ({\em N-average}). In  the same figure a solid line shows this average for $\delta N \approx 2000$. In the present case, such value of $\delta N$ is the required  to obtain a smooth curve.\\
The fluctuations in the fig.~\ref{figure6} are due to the different patterns  of the sequences of $\alpha_{i}'s$ that appear when $N$ is varied.
Therefore, we assume that the  N-average  is equivalent to statistical  averages  over different realizations of the random sequence of $\alpha_{i}'s$.
We define the  average typical current  as  
\begin{equation}
\label{ibarra}
{\hat I}^{2} \equiv \langle I_{typ}^2 \rangle \;
\end{equation}
where the symbol $\langle ..\rangle$ means average over different 
realizations of $\alpha_{i}'s$.\\
$\hat I$ characterizes the magnitude of the  current that is relevant for the experimental investigations \cite{mailly}.\\ 
Before computing $\hat I ^{2}$, we  will give a qualitative explanation of the
origin of the fluctuations in $I_{typ}^2$. 
The total current grows linearly between crossings and therefore its  growth is in direct proportion to the distance between  succesive crossings (see Fig.~\ref{figure3}). There are two extreme  realizations of the sequence of
$\alpha_{i}'s$ that limit the expected behavior of $\hat I^{2}$.\\
Let us first suppose that the crossings conform a regular sequence of equally spaced $\alpha_{i}'s$ (what is  known as a {\bf picket-fence} sequence) and  $\Delta_{i} = \Delta \; \forall i $.
In this case it is easy to verify that  $I_{typ}^{2} \sim {N / n_{c}}^2$. As we have already stated in Sec.~\ref{dc},
$\tilde{n}_{c} \sim \sqrt{N}$ (in this case  $ \tilde{n}_{c} = \; n_{c}$) and therefore $I_{typ}^{2} \;  \sim {N}$.\\
The second realization  corresponds to a sequence of $\alpha_{i}'s$ in which one distance between crossings  is  much larger than the other ones. This resembles the features of the 1D  systems for which in fact, there is only one crossing in the interval  $0 < \alpha < 0.5$. It is straigthforward to show that for the realization  considered $I_{typ}^{2} \; \sim {N}^2$.\\
For the annular billiard, as we have shown in the previous section, the  nearest neighbour spacing distribution between crossings is  Poisson.
Therefore one expects  the actual behavior of $ I_{typ}^{2}$ to be highly 
fluctuating with N around a mean value $\sim N^{\gamma}$ with
$1 < \gamma < 2$.
At the end of the present section  we will  see that the Eq.~(\ref{itypi2}), that depends on the particular realization of the sequence  $\alpha_{i}'s$, fullfills the above expectation. \\

In order to proceed with the computation of  $\hat I ^{2}$,
it is necessary to assume some behavior for the  amplitudes $\Delta_{i}$.
In the  Fig.~\ref{figure3}, it can be seen that the amplitudes of the jumps are not essentially very different on average. 
This fact suggests us to assume 
\begin{equation}
\label{delta}
\Delta_{i}= \Delta = \frac{A_{T}}{2} \; \tilde n_{c} \; ,  
\end{equation}
$\forall i$. This ansatz will be strongly justified by our results. 

From  Eq.~(\ref{itypi2}) and  performing the ensemble average over  $\alpha_{i}'s$  we obtain:
\begin{equation}
\label{ibarra2}
{\hat I}^2 =  2\; \left( \frac{A_{T}^2}{24} - \; \frac{A_{T}}{4} \; \Delta \; {\tilde n}_{c} + \;  A_{T} \; \Delta \; \left( \sum_{j =1}^{{\tilde n}_c} \langle \; 
\alpha_{2j-1}^{2} \rangle \right) + \;  \frac{\Delta^2}{2} \; {\tilde n}_{c}^2 -
\; \Delta ^2 \; \left(  \sum_{j=1}^{{\tilde n}_c} 
(2\; j - 1) \langle \alpha_{2j-1} \rangle \right) \right)\; , 
\end{equation}
where the sums run over the odd values because we have assumed that  the  odd
crossings contribute to the jumps in  the total current (it is worth to note that the election is arbitrary and an equivalent treatment assuming jumps in the even crossings gives the same results).

To evaluate $ \langle \alpha_{2j-1} \rangle $ and 
$\langle \alpha_{2j-1}^{2} \rangle$ we have considered $\alpha_{2j-1}= \; s_{2j-1}$, where  $s_{2j-1}$ is the spacing between  the crossing in $\alpha =0$
and the crossing in $\alpha = \; \alpha_{2j-1}$. Therefore, employing the distributions Eq.~(\ref{poison}) (adequately normalized to nearest neighbour mean spacing  $ d = 1 /(4 \; \tilde{n}_{c})$) we have obtained:
\begin{eqnarray}
\label{prom}
\sum_{j=1}^{{\tilde n}_c} (2\; j - 1) \langle \alpha_{2j-1} \rangle  & = &  d \left(\; \frac{4}{3} \; {\tilde n}_{c}^3 - \; \frac{1}{3} \; {\tilde n}_{c} \right) \; \nonumber \\
\sum_{j=1}^{{\tilde n}_c} \langle \alpha_{2j-1}^2 \rangle  & = &  d^2 \;\left( \frac{4}{3} \; {\tilde n}_{c}^3 + \;  \tilde{n}_{c} ^2 - \frac{1}{3} \; {\tilde n}_{c} \right) \;   \; ,
\end{eqnarray}
here we have employed 
\begin{eqnarray}
\langle \alpha_{2j-1} \rangle  & = &  d \; (2j-1)  \; ,\nonumber  \\
\langle \alpha_{2j-1}^2 \rangle & = &  d^2 \; 2\; j \; (2j -1) \; .
\end{eqnarray}
After replacing Eq.~(\ref{prom}) in Eq.~(\ref{ibarra2}) we finally obtain :
\begin{equation}
\label{ibarra3}
{\hat I}^2 = 2 \left(\frac{ A_{T} ^2}{24} +\; \frac{\Delta^2 \; {\tilde n}_{c}^2}{6} -
\;\frac{A_{T}  \Delta \; {\tilde n}_{c}}{6} + \; \frac{A_{T} \;  \Delta}{16} \; + \frac{ \Delta ^2 }{12} - \; \frac{A_{T} \; \Delta}{48 \; {\tilde n}_{c}} \right) \; .
\end{equation}

The above expression allows us to classify each term according to the dependence on $N$. Taking into account the Eqs.(~\ref{cruces}), (\ref{cruces2}) and (\ref{delta}) and the fact that $A_{T} \propto N$, we can verify that  each of the first three terms in the  Eq.~(\ref{ibarra3}) are of order ${\cal O}(N^2)$ and  that their total contribution vanishes. The fourth  term is of order ${\cal O}(N^{3/2})$ and the two last ones are ${\cal O}(N)$. 
We have obtained for the mean typical current,
\begin{equation}
\label{ibarra4}
{\hat I}^2 =  \frac{ A_{T} ^2}{ 16 \; {\tilde n}_{c}} + \frac{ A_{T}^2}{ 48 \; {{\tilde n}_{c}}^2} \; ,
\end{equation}
and therefore for large $N$, 
\[
{\hat I}^2 \sim  {\cal O} (N^{3/2}) \; .
\]
Eq.~(\ref{ibarra4}) constitutes one of the  main results of the present article.
We have plotted in the Fig.~\ref{figure6}(a), besides the exact numerical results (dots), the N-average with $\delta N = 2000$ (solid line) togheter with its  best power fit $\sim N^{\gamma}$ (dashed line). The resulting value for the exponent is $\gamma =
1.498$ in strong agreement with our theoretical result.  \\
At this point, let us mention that the choice of $\delta N = 2000$ to perform the N-average could resemble somehow ambiguous. However,
if we  perform  an N-average with $\delta N < 2000$, although the resulting curve displays oscillations, the best fit leads to the same scaling with N
(see Fig.~(\ref{figure6})(b)).
It is important to remark that the Eq.~(\ref{itypi2}) can be employed to evaluate the typical persistent current for a given realization of the 
$\alpha_{i}'s$. In particular for the picket-fence sequence (note  the absence of the ensemble average in the following), 
\begin{eqnarray}
\label{pf}
\sum_{j=1}^{{\tilde n}_c} (2\; j - 1) \; \alpha_{2j-1}   & = &  d \left(\; \frac{4}{3} \; {\tilde n}_{c}^3 - \; \frac{1}{3} \; {\tilde n}_{c} \right) \; \nonumber \\
\sum_{j=1}^{{\tilde n}_c} \alpha_{2j-1}^2   & = &  d^2 \; \left(\; \frac{4}{3} \; {\tilde n}_{c}^3 - \; \frac{1}{3} \; {\tilde n}_{c} \right)    \; .
\end{eqnarray}
The last equations lead to a cancellation of all the orders greater  than $\cal{O} (N)$ in the Eq.~(\ref{itypi2}) for $I_{typ}^2$. That is  
\begin{equation}
\label{ipf}
 {I_{typ}^2} = \frac{ A_{T}^2}{ 48 \; {{\tilde n}_{c}}^2} \; .
\end{equation} 
 Equation ~(\ref{ipf}) satisfies $I_{typ}^2 \sim N$  in accordance with our  previous qualitative discussion (see the paragraph below Eq.~(\ref{ibarra})).\\
On the other hand, we expect some other particular realizations of the sequence of $\alpha_{i}'s$ for which there is no cancellation of the terms of order ${\cal O} (N^2)$ in Eq.~(\ref{itypi2}). In this case  $I_{typ}^2 \sim N^2$. \\
We stress that the considered  particular realizations manifest themselves when  $I_{typ}^2$ takes the lowest and highest values respectively (see Fig.~\ref{figure6}(b)). 
 
\section{Concluding Remarks}
\label{cr}
In the present work we have  characterized the behavior of the persistent current for a system of non-interacting electrons confined in a  mesoscopic bidimensional clean cavity  as a function of the number of particles in the system $N$.
We have deduced relevant  properties of the single particle spectrum such as the scaling of the number of open channels $M$ and average number of level crossings ${\tilde n}_{c}$ with $N$.\\
The main goal of this article has been to determine the $N$-average behavior of the typical current through the statistical properties of the degeneracies present in the single particle spectrum as a function of the flux, when many channels are relevant. We can separate the dependence on $N$ of the typical current in two contributions. 
The first one is the average value taking on a range $1<< \delta N < N$ and  constitutes the  smooth contribution.
The second one, depends on  the particular realization  of the sequence of crossings  $\alpha_{i}'s$  (the values of flux where degeneracies in the spectrum occur) and leads to  an extremely fluctuating behavior with $N$. We have related this fact to the statistics of the distribution of crossings which, as we have determined  is Poissonian.
The smooth part of the typical current has been obtained under the assumption that the ensemble average over different realizations of the random sequence of crossings $\alpha_{i}'s$ is equivalent to the N-average.
The obtained dependence on $N$ of this smooth contribution is  ${\hat I}^2 \sim
N^{3/2}$ which is in strong agreement with the  numerical results (see Fig.~\ref{figure6}).
On the other hand we have evaluated the typical current for particular realizations of sequences of  $\alpha_{i}'s$ that lead to  completely different dependence on $N$, namely  $I_{typ}^2 \sim N^{2}$ and $I_{typ}^2 \sim N $.
These realizations are responsible for  the large  oscillations of  $I_{typ}^2$.\\ 
Last but not least, some words of caution are in order in connection with experimental observations of persistent currents. It is necessary 
to emphasize that, owing to the   
extreme sensitivity of the typical current on the particular realization of 
$\alpha_{i}'s$, any  experiment devoted to establish
 the scaling  of the average typical current on N (or on the number of open channels $M$) in mesoscopic 2D cavities would involve  measurements on  the order of a thousand of samples (remember that $\delta N \approx 2000$,
 for the annular geometry).    

\section*{Acknowledgments}
This work was partially supported by UBACYT (TW35), PICT97 03-00050-01015 and CONICET.\\ We would  to thank G. Chiappe for useful discussions.

\newpage

\newpage

\newpage

\newpage

\newpage

\newpage

\end{document}